\documentclass[smallextended, twocolumn]{IEEEtran}  
\usepackage[numbers,sort&compress]{natbib}
\usepackage{tikz}
\usepackage{hyperref}
\usepackage{graphicx}  
\usepackage{qcircuit}

\usepackage{amsmath,amsfonts,bm}
\everymath{\displaystyle} 
\usepackage{colortbl}

\newcommand{\ket}[1]{\ensuremath{\left|#1\right\rangle}} 

\renewcommand{\bf}[1]{\ensuremath{\mathbf{#1}}}

\begin{document}

\title{{A walk through  of time series analysis on quantum computers}


\author{Ammar Daskin\\
\IEEEauthorblockN{
             Dept. of Computer Engineering \\ 
            Istanbul Medeniyet University, Istanbul, Turkey
              \thanks{adaskin25@gmail.com}           
}
}
\date{Received: date / Accepted: date}
}


\maketitle

\begin{abstract}
Because of the rotational components on quantum circuits, some quantum neural networks based on variational circuits
can be considered equivalent to the classical
Fourier networks. 
In addition, they can be used to predict the Fourier
coefficients of continuous functions.
Time series data indicates a state of a variable in time.
Since  some time series data can be also considered as
continuous functions, we can expect quantum machine learning models to 
do many data analysis tasks successfully on time series data.
Therefore, it is important to investigate new quantum
logics for temporal data processing and analyze intrinsic
relationships of data on quantum computers. 

In this paper, we go through the quantum
analogues of classical data preprocessing and forecasting with ARIMA models by using simple quantum operators requiring  a few number of quantum gates.
Then we discuss 
future directions and some of the tools/algorithms that can be used
for temporal data analysis on quantum computers.
\end{abstract}

\section{Introduction}
It is shown that nature itself is explained better in quantum physics than classical one. The idea of quantum computers was put forward by Feynman \cite{feynman1986quantum} and other pioneers \cite{benioff1980computer} so as to simulate and understand intrinsic physical properties of nature in a better way: 
Simulating quantum systems with many particles is difficult on classical computers because the computational complexity in general grows exponentially with the number of involved particles. 
On the other hand, quantum computers are based on qubits implemented by quantum particles which in turn indicates that in principle, we can increase the computational power of quantum computers exponentially by adding  controllable new qubits.
 Therefore, the difficulty of simulating nature for classical computers would become the main advantage of quantum computers. Unfortunately  since the input/output of quantum computers are still in classical ones and zeros, it necessitates a state preparation that involves classical processing of the data and a measurement that requires many repetitions of the same computation. Therefore, the complexity boundaries in classical computation mostly remain the same for quantum computing. 
Nonetheless, it is still possible to design more efficient algorithms by employing quantum entanglement and superposition (quantum parallelism):  Some known examples are Shor’s integer factoring algorithm \cite{shor1994algorithms}, Grover’s search algorithm in an unordered list \cite{grover1997quantum}, and the quantum principal component analysis (PCA) \cite{lloyd2014quantum}.

\subsection{Classical neural networks}
The representation theorem indicates any multivariate continuous function can be represented as a superposition of functions of one variable and addition \cite{kolmogorov1957representation, hecht1987kolmogorov}. 
An artificial neural network can be considered as a composition of simple nonlinear activation function\cite{cybenko1989approximation}:
Let $I^n$ be the n-dimensional unit-cube $[0, 1]^n$ and $C(I^n)$ be the space of continuous functions on $I^n$. In addition, let $\sigma$ be a sigmoidal function. Given any $f \in C(I^n)$ and $\epsilon > 0$, there is a $G$ of the following form:
\begin{equation}
G(x) = \sum_{j=1}^{n} \alpha_j \sigma(w_j^Tx+\theta_j)
\end{equation}
for which $|G(x)-f(x)|<\epsilon$. 

Gallant et al. \citep{gallant1988there} showed a single feed-forward neural network where the output is squashed by using a cosine activation function: i.e. by considering $\sigma$ as cosine, sine, or exponential (i.e. in the form $e^{iw_j^Tx+\theta_j}$) functions, one can approximate the Fourier transform of a continuous function. These networks are often called Fourier neural networks (FNN).
On the other hand, Fourier analysis of data also sheds lights on the generalization performances of deep neural networks\cite{xu2020fprinciple}.

\subsection{Time series problems}
Temporal data indicates a state of a variable in time, which could be collected through a wide variety of monitoring devices. 
Examples include financial market data; sensor data which may indicate temperature, humidity and so on;
and medical data collected by different monitoring medical devices. Depending on the collection method, the data can be continuous or discrete. 
Continuous temporal data sets are generally referred to as times series data. \cite{aggarwal2015data}
The data mining tasks on time series data include clustering, classification, and forecasting: In general it is required to specialize the known algorithms in data analysis in order to use in the analysis of time series data: e.g. see the classification 
with deep learning for time series data \cite{ismail2019deep} and the distance based time series classification \cite{abanda2019review}.
It is shown that time series data can be also modeled by using Fourier deep neural networks \cite{gashler2016modeling}, which is beriefly explained above.

\subsection{Quantum machine learning}
Since the quantum machine learning \cite{dunjko2020non,dunjko2018machine} is a very new field, there are only a small number of partial studies which can be found for the analysis of time series data on quantum computers: the time series data analysis is done by using a quantum approach based on the Schr{\"o}dinger's equation \cite{singh2018quantum}. 
In a very recent paper, the quantum machine learning approaches are investigated for the problems in finance \cite{emmanoulopoulos2022quantum}. 
It is shown that the variational quantum machine learning models can be used to approximate Fourier transforms of the continuous functions \cite{schuld2021effect}.
It is also shown that the continuous variable (CV) quantum neural networks can be used for curve fitting and function approximations\cite{killoran2019continuous}. 
It is still important to study and build generic separate models and packages for temporal data analysis.
 
Quantum circuits are in general based on quantum gates described by some form of rotations in the Hilbert space. For instance, the rotation gates for one-qubit can be defined in the following forms;
\begin{equation}
R_x(2\theta) = \left(\begin{matrix}
\cos(\theta) & -i\sin(\theta)\\
-i\sin(\theta) & \cos(\theta)
\end{matrix}\right),
\end{equation}

\begin{equation}
R_y(2\theta) = \left(\begin{matrix}
\cos(\theta) & -\sin(\theta)\\
\sin(\theta) & \cos(\theta)
\end{matrix}\right),
\end{equation}

\begin{equation}
\text{and } R_z(2\theta) = \left(\begin{matrix}
e^{-i\theta} & 0\\
0 & e^{i\theta}
\end{matrix}\right).
\end{equation}
Because of these rotational components, the quantum circuits can in theory provide a natural equivalent to the classical Fourier networks (e.g., see our previous paper\cite{daskin2018simple} which gives an example quantum neural net with periodic activation function.). In addition, they can be used to predict Fourier coefficients of continuous functions as done in Ref.\cite{killoran2019continuous}

\subsection{Quantum temporal data analysis}
\begin{figure*}
	\includegraphics[width=\textwidth]{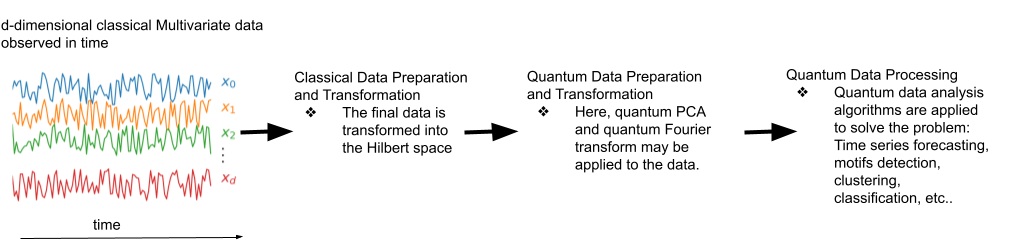}
	\caption{	\label{FigGeneralApproach} General picture for processing temporal data on quantum computers (the measurement step is not shown.)}
\end{figure*}

Since some time series data can also be considered as continuous functions, quantum machine learning models can successfully do many data analysis tasks on the time series data. Therefore, it is important to investigate new quantum logics for temporal data processing and  analyze intrinsic relationships of data on quantum computers. 
The processing data on quantum computers in general requires the followings:
\begin{itemize}
\item classical preprocessing and quantum state preparation, \item quantum preprocessing, \item quantum processing of data, \item quantum measurement, \item and analyses of the measurement results on classical computers
\end{itemize} 
In general, all or some of these steps are repeated until a desired result/model is obtained in data analysis. 
For continuous temporal data, these steps are visualized in Fig. \ref{FigGeneralApproach}. 

In the following sections,  we go through the quantum analogues of classical data preprocessing and forecasting. 
Then we discuss some of the tools/algorithms that can be used for temporal data analyses on quantum computers and discuss future directions.

\section{Preparing data for quantum algorithms}
The time series data is generally given in contextual representations where attributes describe the context and behavioral values at any reference points \cite{aggarwal2015data}. 
If more than one behavioral attribute is associated with the series, then the data is referred as the multivariate time series: i.e. A time series data with dimension $d$ and length $n$. The data is formed by a set of the values received at the time stamp $t_i$. In the $i$th time stamp, the received data is a vector of dimension $d$: $\bf{y_i} = (y_i^1, \dots, y_i^d)$.

In classical data analyses, the time series data is taken under preprocessing operations: e.g., the missing values may be handled,the data may be normalized, smoothed, or scaled. Then, a variety of methods are used to transform and reduce the dimension of the data.
After these steps, the standard data analysis tasks such as classification and clustering are also applicable to time series data.
The classification and clustering tasks involve determining the correlation across the series and classifying the series based on their shapes. These can be offline or real time. And class labels may need to be assigned to the individual time stamps (point labels) or the whole series.

For quantum computers, the algorithms either use the data after some classical preprocessing, or as done in PCA \cite{lloyd2014quantum}, it assumes the data is processed and stored in quantum RAM \cite{giovannetti2008quantum}. In the former case, the data is prepared as a quantum state by a state preparation circuit that converts a standard initial state:e.g. $\ket{0}$, into the state that represents the data $\ket{x}$. If the length of \ket{x} is $n$, then this kind of state preparation requires a circuit with $O(n)$ number of controlled-NOT quantum gates in general \cite{nielsen2002quantum}.
In some cases, the data is also used as parameters that determine the rotation angle of the quantum gates(e.g.  in \cite{perez2020data}, the data is used with parameters to define quantum rotation gates. The learning task is done by reusing the data and adjusting the value of the parameters for quantum gates.).
After the mapping, \ket{x} is a vector defined in the Hilbert space. It is shown in Ref.\cite{schuld2019quantum} that the Hilbert space can be considered as a kernel space used in classical machine learning to ease the training process. Therefore,  the training of a learning model can be achieved also in the Hilbert space.
Although this solves the representability of the data in the Hilbert space, it is still necessary to design efficient data mapping methods without disrupting the representability of the data (e.g. for MNIST data, it is reduced to 8 qubits by using classical PCA \cite{grant2018hierarchical} which reduces the training results to as low as 70\%. ).

Consequently, the representation is highly dependent on the quantum algorithm to be used in the data processing step. However, 
in most cases, since the time series data is  multivariate in the form $\bf{y_i} = (y_i^1, \dots, y_i^d)$,  the quantum superposition approach can be used to efficiently represent the data on quantum computers. This in general may reduce the $d\times n$ data into an $n$-dimensional quantum state:
\begin{equation}
    \ket{\bf y} = \frac{1}{\sqrt{n}}\sum_i^n \ket{\bf{y_i}}
\end{equation}
Alternatively, one can build the following quantum state from the input data:
\begin{equation}
       \ket{\bf y} = \left( \begin{matrix} \bf{y_1} \\ \bf{y_2}\\ \vdots \\ \bf{y_n} \end{matrix}\right)
\end{equation}
Here, the dimension of the quantum state is $\log(nd)$.
Another alternative can be the following:
\begin{equation}
    \ket{\bf y} = \ket{\bf y_1}\ket{\bf y_2}\dots \ket{\bf y_n}.
\end{equation}
Although the required number of qubit which is $n\log(d)$ is much more than the first two methods, this representation may allow us to apply quantum operations to individual set of values more easily. However, since the required number of qubits depend on $n$, it may not be possible on current quantum computers to use this representation for the series with large $n$.

\section{Quantum preprocessing of temporal data}
Below we show how to implement quantum analogues of some of the famous classical preprocessing techniques: the exponential smoothing, binning and  data transformation. Note that in most cases, we go through the steps for univariate data.
\subsection{Exponential smoothing}
In the exponential smoothing, the smoothed value $y'_i$ is obtained through the combination of $y_i$ and the pervious values with the ratios determined by the decay factor $\alpha$:
\begin{equation}
	y'_i = \alpha y_i + (1-\alpha)  y'_{i-1}.
\end{equation}
The recursive substitution in the above equation yields:
\begin{equation}
y'_i = (1-\alpha)^i y'_0 +\alpha \sum_{j=1}^{i} y_j (1-\alpha)^{i-j}.
\end{equation}
In the analyses, the choice of $\alpha$ determines the impact of the previous data points. 

In quantum analyses, we can construct a similar exponential smoothing technique:

\begin{enumerate}
\item   
Different from the classical analysis, \ket {\bf{y_i} },  the data observed at the $i$th time stamp,  can be a single data point or  multiple data points observed on $d$ dimension. Therefore, as a physical qubit, this state can be  implemented as a $d$-dimensional quantum system with $O(\log(d))$ qubits or a $d$-level qubit. 
\item 
The new observed data at time stamp $(i+1)$ is prepared as a quantum state $\ket {\bf{y_{i+1}} }$ on different qubits: i.e., \ket {\bf{y_{i+1}} }\ket {\bf{y_{i}} }).
Or we can create the following quantum state by using a control qubit and data qubit. 
\begin{equation}
    \left(\begin{matrix}
    \ket {\bf{y_{i+1}} }\\\ket {\bf{y_{i}} })
    \end{matrix}
    \right)
\end{equation}
Note that this state can be formed with the help of control qubit  by applying the operators related to \ket {\bf{y_{i+1}} } and \ket {\bf{y_{i}} }) to the data qubit in \ket{\bf 0} state. 
\item From the above state, a quantum entangling gate (e.g. the Hadamard gate) can be used to generate the following recurrence:
 	\begin{equation}
 	\begin{split}
 	     		\ket{\bf{y'_{i+1}}}  =& \sqrt{\alpha} \left( 
 		\ket {\bf{y_{i+1}} } + \ket {\bf{y'_{i}} }\right) 
 		+ \sqrt{1- \alpha} \left(\ket {\bf{y_{i+1}} } - \ket {\bf{y'_{i}} }\right)\\
 			\\ 
 	= &	a \ket {\bf{y_{i+1}} } 		+ b\ket {\bf{y'_{i}} }	
 	\end{split}
 	\end{equation}
where $a$ $=$  $\left(\sqrt{\alpha} + \sqrt{1- \alpha} \right)$ and $b$ $=$ $\left(\sqrt{\alpha}-\sqrt{1- \alpha} \right).$  	
    	In the case of the Hadamard gates, $\alpha$ is some power of $1/2$.
Using the above recursion, if we substitute \ket{ \bf{y'_{i}}} into the above, \ket{\bf{y'_{i+1}}} becomes:
\begin{equation}
\begin{split}
\ket{\bf{y'_{i+1}}} = & 
 		a  \ket {\bf{y_{i+1}} } 	
+
b 
\left(
a \ket {\bf{y_{i}} } + b \ket {\bf{y'_{i-1}}} 
 \right) \\   	 
 =& a\ket {\bf{y_{i+1}} } + ab \ket{\bf{y_{i}} } +ab^2 \ket {\bf{y_{i-1}} } \dots \\
 =& ab^{i+1} \ket {\bf{y'_{0}} } 
 + \sum_{k=1}^{i} a b^{i-k} \ket{\bf{y_{k+1}} }, 
\end{split}
\end{equation}
If we choose an $\alpha$ which makes $b < 1$, we can have the quantum equivalent of the classical exponential smoothing.
In this case, the choice of $\alpha$ regulates the decay of the terms through $b$; therefore, we can call $b$ as the decay factor that determines the importance of the recent data points in any analysis using $\ket{\bf{y'_{i}}}$.
 
\end{enumerate}
\subsection{Binning and moving average smoothing}
The classical moving average is a special case of the binning method where the date is divided into equally spaced time intervals of size $k$: $[ t_1, t_k], [t_{k+1}, t_{2k}]$.
Then, the data points are assigned into the corresponding bins and the mean values of the bins are used for the analyses:
\begin{equation}
y'_{i+1} = \frac{\sum_{r=1}^{k}y_{i.k+r}}{k}	
\end{equation}
Here, the number of the data points are reduced by a factor of $k$.

Utilizing the Hadamard gates on certain qubits, one can obtain the average of the neighboring states on certain quantum states. 
After finding the average, we can also disregard the rest of the state by collapsing quantum state onto the interested part of the system:  One example could be as follows:
\begin{itemize}
\item Let $\ket{\bf{y_{i}}} = \left(\begin{matrix}
y_{i1}\\
\vdots\\
y_{id}
\end{matrix}\right)$.
If we multiply this quantum state $U_{bin}= I^{\otimes \log_2(d/k)}\otimes H^{\otimes \log_2 k}$, where $H$ is the Hadamard gate:
\begin{equation}
H = 1/\sqrt{2}\left(\begin{matrix}
1&1\\
1&-1
\end{matrix}\right)
\end{equation}
If we apply $U_{bin}$ to the quantum state $\ket{\bf{y_i}}$, then when the first $\log_2k$ qubits in \ket{0\dots0} state, we would have an equivalent $\bf{y'_{i+1}}$ which could be considered reduced data points.
\end{itemize}

The classical moving average smoothing uses overlapping bins which can be easily done on quantum case by utilizing a quantum operator of the form: $H^{\otimes \log_2 k}\otimes I^{\otimes \log_2(d/k)}$.

 \subsection{Data transformation }
 In classical data analyses, often the temporal data is made smaller or discreticezed through data transformation methods such as discrete wavelet and Fourier transforms (DWT and DFT respectively).
 
 Similar to the Fourier transform, DWT represents a data (function) in terms of an orthonormal basis. One of the simplest wavelet transform is the Haar transform which can be described a  orthogonal matrix: e.g., $4\times 4$ Haar transform is
\begin{equation}
H_{4}={\frac {1}{2}}{\begin{bmatrix}1&1&1&1\\1&1&-1&-1\\{\sqrt {2}}&-{\sqrt {2}}&0&0\\0&0&{\sqrt {2}}&-{\sqrt {2}}\end{bmatrix}}
\end{equation} 
  Similarly to Walsh-Hadamard transform, the unnormalized Haar transform matrix can be obtained by the following recurrence:
 \begin{equation}
H_{{2N}}={\begin{bmatrix}H_{{N}}\otimes [1,1]\\I_{{N}}\otimes [1,-1]\end{bmatrix}}.
 \end{equation}
 The transformation on an input vector $\bf{y'}$ is described by $\bf{y'} = H_N \bf{x}$. From this product, the rows of the Haar matrix allow us to draw different properties of the temporal data: For instance, for $H_4$, the first row takes the average of the input vector, while the second row measures the low frequencies. The rest of the rows measures different parts of the input vector.

In the wavelet transform, $\bf{y}$ can be represented with the basis vectors that form $H_n$:
\begin{equation}
\bf{y'} = \sum_{i=1}^{q} a_i \bf{w_i},
\end{equation}
where $\bf{w_i}$ is the $i$th basis vector and $a_i$ is the normalized coefficient. The dimension reduction is done by disregarding the lowest $a_i$s.

In quantum computing, since the Haar matrix is orthogonal and can be built from the recurrence that is described by the Kronecker product, it can be implemented as a quantum gate in a similar fashion to the Hadamard gate. Therefore, given \ket{\bf{y}}, one can easily obtain $\ket{\bf{y'}}$. The elimination of lowest parts of $\ket{\bf{y'}}$ can be done through either sampling or measuring a few qubits. This leads to a problem of finding the best qubits that gives the maximum reduction with the minimum information loss.

 \subsubsection{Quantum Fourier transform (QFT)}
The discrete Fourier transform (DFT) can be used to transform a data vector into a linear combination of sinusoidal series. The similarity of two vectors can be measured by taking the Euclidean distance of the Fourier coefficients.
In quantum computing, DFT-which has the complexity bound $O(nlog n )$ for $n$-dimensional vector-can be implemented with $O(logn)$ computational steps.
It provides the main speedup of many quantum algorithms such as integer factoring.
The quantum Fourier transform for the data $\ket{\bf{y}}$ can be obtained by applying the operation $QFT$. If we have two data vectors $\bf{y_1}$ and $\bf{y_2}$ of the same dimension $n$ (we assume this for convenience and brevity  in notations, it can be easily generalized to any dimension.), we first construct the following quantum state:
\begin{equation}
\ket{\bf{y}} = \left(\begin{matrix}
\bf{y_1}\\
\bf{y_2}
\end{matrix}\right)
\end{equation}  
Then we apply the quantum Fourier transform to each part of the vectors:
\begin{equation}
\left(\begin{matrix}
QFT& \\
&QFT
\end{matrix}\right)
\ket{\bf{y}} = \left(\begin{matrix}
QFT\bf{y_1}\\
QFT\bf{y_2}
\end{matrix}\right)
\end{equation}
Here note that 
\begin{equation}
\left(\begin{matrix}
QFT& \\
&QFT
\end{matrix}\right) = I \otimes QFT,
\end{equation}
where $I$ is a $2\times 2$ identity matrix. We then apply the Hadamard gate to the first qubit:
\begin{equation}
\left(H \otimes I^{\otimes log(n)} \right) \left(\begin{matrix}
QFT\bf{y_1}\\
QFT\bf{y_2}
\end{matrix}\right) = \frac{1}{\sqrt{2}}\left(\begin{matrix}
QFT\bf{y_1}+QFT\bf{y_2}\\
QFT\bf{y_1}-QFT\bf{y_2}
\end{matrix}\right)
\end{equation} 
We can also start with a superposition of the input state:
\begin{equation}
\ket{\bf{y}} = \frac{1}{\sqrt{2}}\left(\ket{\bf{y_1}} - \ket{\bf{y_2}}\right).
\end{equation}  
Then we apply $QFT$:
\begin{equation}
QFT\ket{\bf{y}}= \frac{1}{\sqrt{2}}\left(QFT\ket{\bf{y_1}} - QFT\ket{\bf{y_2}}\right).
\end{equation} 
This also generates a kind of the desired distance. 

As mentioned before, we can also use a separate register for each data point at different time: e.g.,  the quantum state \ket{\bf{y_1}}\ket{\bf{y_2}} is for the data measured at time t=1 and t=2. In this case we apply $QFT$ to both registers and then use the SWAP test to create the distance measure between data points. 
\begin{equation}
   \left(
QFT\otimes QFT\right) \ket{\bf{y_1}}\ket{\bf{y_2}}= QFT\ket{\bf{y_2}}QFT\ket{\bf{y_2}}
\end{equation}
In all cases, the Euclidean distance can be obtained by the measurement results.
However, if $d$ number of $\bf{y_i}$ are given as classical input,  the first approach requires $O(log(dn))$ number of qubits and the third approach requires $O(dlog(n))$ since a separate register is used for each $\bf{y_i}$.
Since the second approach uses the superpositioned state, the number of qubits is $O(\log(n))$ which is less than the first and third approaches. 
 On the other hand,   the classical processing time is almost the same for all approaches and is $O(nd)$ since it requires linear scanning of the every data points. 
However, if $\ket{\bf{y_i}}$ are stored on a random accessed quantum memory (qRAM), then all approaches  requires  only $O(d)$ number of queries to qRAM to load the data into quantum registers.

\section{Time series forecasting}
If the statistical parameters of time series such as the mean and variance do not change with time, it is called stationary: i.e. the probabilistic distribution of the parameters in any time interval is the same as or very close to the time interval found by shifting from this interval. And non-stationary if the distributions change with time, which is mostly the case when we deal with real world data: e.g., if we look at the number of daily COVID cases, we see that the daily cases changes based on the season and different factors  for the past few years.  
A nonstationary series can be made stationary by different operations on the times series data. 
\subsection{Differencing}
Time series can be converted by considering differences between the data $y_i$ at time $t_i$ and the data $y_{i-1}$ at the previous time $t_{i-1}$. This operation is called differencing and defined as follows:

\begin{equation}
y'_i = y_i - y_{i-1}.
\end{equation}

If the data is still not stationary, then a second order differencing also can be applied:
 \begin{equation}
 y''_{i} = y'_i+ y'_{i-1}= y_{i}-2y_{i-1}+y_{i-2}
 \end{equation}
To implement this as a quantum operation, first we observe that the closer $\ket{y'}$ to the equal superposition state, $\ket{H} = H^{\otimes n}\ket{0}$, the more stationary it is.
We can measure the closeness of the two quantum state by the swap test\cite{buhrman2001quantum} or by some statistical sampling from the quantum state(i.e., measuring some qubits to obtain probabilities of 0 and 1. The same or close values indicate an equal superposition state). 

To  construct $\ket{y'}$ first we use an auxiliary qubit in \ket{1}, then apply a Hadamard gate to this qubit:
\begin{equation}
\left(H\ket{1}\right)\ket{y} =\frac{1}{\sqrt{2}}\left(\begin{matrix}
\ket{y}\\
-\ket{y}
\end{matrix}\right)
\end{equation} 
Then we apply laddered Toffoli operations 
(similar to a bit-adder: See quantum adders in Ref.\cite{nielsen2002quantum}) to the second part of the state to convert into the following:
\begin{equation}
\frac{1}{\sqrt{2}}\left(\begin{matrix}
\ket{y}\\
-U_{adder}\ket{y}
\end{matrix}\right)
=\left(\begin{matrix}
y_0\\
y_1\\
\vdots\\ 
y_{n-1}\\ 
-y_{n-1}\\
-y_0\\
\vdots\\
-y_{n-2}
\end{matrix}\right) 
\end{equation}
Then, we apply a Hadamard gate again to the first qubit to obtain the following
\begin{equation}
\label{Eqqstate}
\left(H\otimes I^{\otimes n-1}\right)\frac{1}{\sqrt{2}}\left(\begin{matrix}
y_0\\
y_1\\
\vdots\\ 
y_{n-1}\\ 
-y_{n-1}\\
-y_0\\
\vdots\\
-y_{n-2}
\end{matrix}\right) 
= \frac{1}{2}\left(\begin{matrix}
y_0-y_{n-1}\\
y_1-y_{0}\\
\vdots\\ 
y_{n-1}-y_{n-2}\\ 
y_0+y_{n-1}\\
y_1+y_{0}\\
\vdots\\ 
y_{n-1}+y_{n-2}\\ 
\end{matrix}\right) 
\end{equation}
Note that by applying Hadamard gates on different qubits and some swap operations we can easily built $\ket{y''}$ or any high order differencing from the above state. 
For instance, one can obtain some seasonal differencing as follows:
For a natural number $s<i$, let the seasonal difference be defined as
\begin{equation}
y'_i = y_i-y_{i-s}.
\end{equation} 
In Eq.\eqref{Eqqstate}; instead of applying the Hadamard gate to the first qubit, if we apply to the second or any other qubit based on the value of $s$, we basically construct the equivalent quantum state for seasonal difference.
Also note that the desired quantum state is the half part of the above state where the first qubit is in \ket{0} state. 
In the further computations, one can either collapse the state to the desired part by measuring the first qubit in \ket{0} state\footnote{For positive $\bf{y}$,   we can also apply an oblivious-ampltidue amplification \cite{berry2017exponential} to the first qubit or consider a different order of differencing}.

\subsection{Forecasting Models}
Differencing mean the time series is drifting in time; therefore, it can be modeled by simply the following:
\begin{equation}
y_{i+1} = y_i +c +e_{i+1},
\end{equation} 
where $y_{i+1}$ is the predicted value,  $e_{i+1}$ represents the white noise with zero mean, and $c$ is the time drift. 

Forecasting on time series data can be made through autoregressive 
models.  The forecasting with the univariate time series data is generally based on autoregression models where the output value at time $t$, $y_t$, is determined as a linear combination of the values preceding  window of some length $p$ (AR($p$) model)\cite{aggarwal2015data}:
\begin{equation}
y_t = \sum_{i=1}^{p} a_i \cdot y_{t-i}+c+\epsilon_t,
\end{equation}
where $c$ and the coefficients $a_1 \dots a_p$ are learned through a learning process.

In the moving average model (MA($q$)), the behavioral attribute value at any timestamp is determined from the unexpected historical variations in the time series data (shocks):
\begin{equation}
y_t = \sum_{i=1}^{q} b_i \cdot \epsilon_{t-i}+c+\epsilon_t,
\end{equation}
where $c$ is the mean value and the coefficients $b_i$s are learned from the data.
A more powerful and general model can be obtained by combining the two aforementioned models (ARMA($p,q$)):
\begin{equation}
\label{Eq:ARMA}
y_t = \sum_{i=1}^{p} a_i \cdot y_{t-i}+c+\epsilon_t + \sum_{i=1}^{q} b_i \cdot \epsilon_{t-i}+c+\epsilon_t.
\end{equation}

Autoregressive moving average models (ARMA) are used best with the stationary data (when the mean, variance, and autocorrelation does not change in time). The non-stationary data are handled by integrating a series of differences into the model (ARIMA($p,q,q$)). For instance with the first-order difference ($d=1$) the model can be written as:
\begin{equation}
\label{Eq:ARIMA}
y'_t = \sum_{i=1}^{p} a_i \cdot y'_{t-i}+c+\epsilon_t + \sum_{i=1}^{q} b_i \cdot \epsilon_{t-i}+c+\epsilon_t.
\end{equation}
We can easily generate the above by using a variational quantum circuit including two and single qubit gates (as done in many of quantum machine learning models) and apply to  the quantum state prepared in one of the following forms: 
\begin{equation}
\ket{\bf y}\ket{\bf\epsilon}, \left(\theta_y \ket{\bf y}+\theta_{\epsilon}\ket{\bf\epsilon}\right), \text{ or } \left(\begin{matrix} \ket{\bf y}\\ \ket{\bf\epsilon}\end{matrix}\right) . 
\end{equation}
Note that \ket{\bf y} may also represent the state after differencing. 
Also note that the angle values of the gates in the circuit are related to the parameters $\bf{a}$ and $\bf{b}$ so that we can obtain the parameterized model in Eq.\eqref{Eq:ARIMA} or any of the previous models.
Furthermore, note that following the Ref.\cite{daskin2012universal} where it is shown how to map the Schmidt decomposition of a vector into a quantum circuit,  one can also prepare a quantum operator whose first row is the parameters $\bf{a}$ and $\bf{b}$. 
\begin{align}
    \left(\begin{matrix}
    a_1 & \dots & a_p &b_1\dots&b_q\\
    \bullet & \dots & \bullet&\dots&\bullet\\
    \vdots & \vdots & \vdots&\vdots&\vdots\\
    \end{matrix}\right)
\end{align}
If the dimensions of $\bf a$ and $\bf b$ are $k$, this operator can be generated by using $O(k)$ number of quantum operations.
We map the parameters to the angles of the rotation gates. Then, we adjust the rotation gate parameters by using a classical optimization technique as done in the variational solvers.

Since the value of $p$ determines the seasonality (period) in the time series. Therefore, ARIMA model indicates the auto-correlation between a value and its neighbors.
 In some ARIMA like models, $p$ can be considered as the period of the time series. 
 On quantum computers the value of $p$  can be found along with the other optimization parameters  by using a classical optimization.
 In addition, since it is the period of the data, one can also adapt quantum phase estimation algorithm to find this period as done in Shor's factoring algorithm. 
\subsection{Forecasting Models As Circulant Matrix Operators}
If the equation for $y_t$ in any of the models is stacked up for the values $t = 0, 1, \dots, T-1$, the above generic models can be rewritten as a system of equations constructed by circulant matrices \cite{pollock2002circulant}. 
The circulant matrices can be implemented very efficiently on quantum computers since their eigenspace is formed by the Fourier transform (see Ref.\cite{daskin2022circulant} for circuit implementation of these matrices). Therefore, one can use those matrices in conjunction with the variational quantum circuits to optimize the parameters of the forecasting models.

\section{Discussion and Future Directions}
In this paper, we  go through data processing and forecasting with ARIMA models and show how to implement them on quantum computers.
As future direction,  more example data analysis are needed 
to investigate algorithms and methods for preprocessing and transformation of time series data into the Hilbert space in which quantum computers work better than classical computers; and investigate quantum machine learning methods and algorithms for univariate and/or multivariate time series data to forecast, classify and cluster the data. 
In addition, for time series data, there are many classical packages such as Facebook's- Prophet \cite{facebook2022} and ARIMA (Autoregressive Integrated Moving Average) models. There is also need to design a quantum packages-that can be used with the current quantum libraries such as IBM-QISKIT\cite{qiskit2022}-for time series data to forecast and do other tasks.

\subsection{Data representation and Measurement}
Representation of data on quantum computers may not be without noise. Therefore, the small changes in any feature of data may not be observed on quantum state by using simple measurements.
If one qubit is used for one feature, then we expect any change in the input data to impact the measurement statistics of a qubit.  
However, representing whole feature vector as a quantum state with fewer qubits enforces an automatic dimension reduction in the measurement and hence may cause a significant data loss that impedes the prediction results.

\subsection{Quantum optimization for autoregressive models}
Note that autoregressive models can be also combined with quantum neural networks as done with the classical neural networks \cite{zhang2003time}. 
Also note that autoregressive models can be formulated as combinatorial optimization problems in the framework of graph learning \cite{peng2021graph}. 
In addition, it is shown that quadratic unconstrained optimization formulation can be used for forecasting in finance \cite{orus2019forecasting}. 
With the help of these and other similar studies,  autoregressive models can be formulated as combinatorial optimization in the form of quadratic unconstrained binary optimization. Therefore, the quantum optimization approaches such as adiabatic quantum computation \cite{farhi2000quantum}, quantum unconstrained bounded optimization algorithm \cite{farhi2014quantum}, quantum power iteration \cite{daskin2021combinatorial}  can be used for these models.


\bibliographystyle{IEEEtran}
\bibliography{main}

\end{document}